\documentclass[aps,pra,preprint,amsmath,amssymb,showpacs]{revtex4}
\usepackage{color}
\usepackage{epsf}
\usepackage{epsfig}
\usepackage{graphicx}
\usepackage{latexsym}
\usepackage{bm}
\usepackage{amssymb}
\usepackage[english]{babel}

\begin{document}


\title {Comment on ``Dynamics of transfer ionization in fast ion-atom collisions''}

\author{Yu.\,V.\,Popov}
\email{popov@srd.sinp.msu.ru}
\affiliation{Skobeltsyn Institute of Nuclear Physics, Lomonosov
Moscow State University, Moscow 119991, Russia}
\author{V.\,L.\,Shablov}
\affiliation{Obninsk Institute for Nuclear Power Engineering,
National Research Nuclear University, Obninsk, Kaluga region
249040, Russia}
\author{K.\,A.\,Kouzakov}
\affiliation{Department of Nuclear Physics and Quantum Theory of
Collisions, Faculty of Physics, Lomonosov Moscow State University,
Moscow 119991, Russia}
\affiliation{Skobeltsyn Institute of Nuclear Physics, Lomonosov
Moscow State University, Moscow 119991, Russia}
\author{A.\,G.\,Galstyan}
\affiliation{Department of Nuclear Physics and Quantum Theory of
Collisions, Faculty of Physics, Lomonosov Moscow State University,
Moscow 119991, Russia}

\begin{abstract}
We inspect the first-order electron-electron capture scenario for transfer
ionization that has been recently formulated by Voitkiv \emph{et al.}
(Phys. Rev. A \textbf{86}, 012709 (2012) and references therein). Using
the multichannel scattering theory for many-body systems with Coulomb
interactions, we show that this scenario is just a part of the
well-studied Oppenheimer-Brinkmann-Kramers approximation. Accurate
numerical calculations in this approximation for the proton-helium
transfer ionization reaction exhibit no appreciable manifestation of the
claimed mechanism.
\end{abstract}


\maketitle

\section{Introduction \label{introduction}}

Recently Voitkiv \emph{et al}. published a series of
papers~\cite{Voitkiv_PRL,Voitkiv_JPB,Voitkiv_PRA} putting forth a new
first-order capture mechanism that can be called electron-electron Auger
(or $ee$-Auger)~\cite{Voitkiv_PRA}. According to this mechanism, the
electron undergoes a nonradiative transition from the atomic state to the
bound state of the projectile, transferring the energy excess to the
another atomic electron which is emitted from the atom. This scenario
resembles a kind of Auger decay and to be contrasted with the first-order
radiative capture~\cite{Briggs1974} which is accompanied by emission of a
photon instead of an electron. A clear signature of the $ee$-Auger
mechanism, according to Voitkiv \emph{et al}., is emission of the electron
in the direction opposite to the projectile motion (in the rest frame of
the atom).

A quantum mechanical explanation how a target electron can be captured
into a bound state of a fast moving projectile (proton) was given by
Oppenheimer, Brinkmann and Kramers (OBK)~\cite{OBK}. In the OBK scenario
the electron transfer proceeds via an overlap of initial and final wave
functions of the projectile-target system. This so-called kinematical
capture relies strongly on the radial and angular electron correlations in
the target, if we consider transfer excitation (TE) and transfer
ionization (TI) processes. It must be noted that in quantum mechanics the
transition of an electron to the projectile bound state can be
nonradiative, and the energy excess can be carried away by a third body
that participates in the reaction. It is the first-order, Born term.

In quantum mechanics fast processes are usually treated within Born
approximations. This framework is directly applicable in the case of
two-body scattering, but requires additional careful considerations in the
many-body case. The situation of  particular importance is when the
entrance channel of the reaction is different from its exit channel, for
example, as it is in capture processes. Within the multichannel scattering
theory, the OBK mechanism can be attributed to the first Born
approximation (FBA), whereas the $Ne$- \cite{Thomas1927} and $ee$-
\cite{Briggs1979} Thomas mechanisms can be described using the second Born
approximation (SBA). Any Born approximation is a sum of matrix elements.
Each of them corresponds to a particular interaction that enters a total
perturbation potential. For example, the OBK matrix element is one of the
three FBA terms (see below). The SBA contains twelve different terms, and
only two of them correspond to the $Ne$- and $ee$-Thomas mechanisms.

In this Comment, we examine the mechanism suggested by Voitkiv \emph{et
al}. on the basis of consistent multichannel scattering theory. We show
that, in contrast to the claim of Ref.~\cite{Voitkiv_PRL}, it is not new
and previously undiscussed. Namely, it is just a part of the usual
kinematic capture in the OBK approximation, and the correlated
electron-electron emission is nothing else but a result of the
electron-electron correlations in the target atom. Moreover, the main
formula, employed by Voitkiv \emph{et al}. for the transition amplitude,
contains apparent flaws.

Voitkiv \emph{et al}. use in their
works~\cite{Voitkiv_PRL,Voitkiv_JPB,Voitkiv_PRA} a time-dependent
approach. In this Comment, we consider a time-independent formulation,
noting that both treatments are equivalent at high projectile
velocities~\cite{Belkic}. Atomic units (a.u., $\hbar = e = m_e = 1$) are
used throughout unless otherwise specified.

\section{Elements of multichannel scattering theory \label{theory_s}}
In this section we remind basic formulas of quantum scattering theory for
many-body systems. More mathematical details, particularly for the case
involving charged fragments, one can find in the review
articles~\cite{Belkic,Shablov2010}. A set of relative momenta defining
motion of $n_{\alpha}$ fragments colliding in the asymptotic channel is
denoted by $\vec p_\alpha$. In turn, the ket vector
$|\phi_{\alpha}\rangle$ stands for a product of bound (spectral) state
wave functions, which define the channel $\alpha$. Hence, the ket vector
$|\phi_{\alpha},\vec p_\alpha\rangle$ is the eigenfunction of the
asymptotic hamiltonian $H_\alpha$: $(E-H_\alpha)|\phi_{\alpha},\vec
p_\alpha\rangle=0$. The total hamiltonian is
$H=H_\alpha+V_\alpha=H_\beta+V_\beta$, where $V_\alpha\ (V_\beta)$ is a
sum of two-body interaction potentials, which we consider as perturbation,
and they define the terms of the Born series: FBA, SBA, and so on.

The amplitude of the transition from the channel $\alpha$ to the
channel $\beta$ can be presented using two forms. These are the
{\it post-}form
$$
T_{\beta\alpha}(E)= \langle\phi_{\beta},\vec
p_\beta|V_{\beta}|\Psi^+_{\alpha}(\vec
p_\alpha)\rangle, \eqno (1)
$$
and the {\it prior-}form
$$
{\tilde T}_{\beta\alpha}(E)= \langle\Psi^-_{\beta}(\vec
p_\beta)|V_{\alpha}|\phi_{\alpha},\vec p_\alpha\rangle, \eqno (2)
$$
where $(E-H)|\Psi^{\pm}_{\alpha(\beta)}(\vec p_{\alpha(\beta)})\rangle=0$.
It is straightforward to show that (see, for instance,
Refs.~\cite{Taylor,Joachain})
$$
T_{\beta\alpha}(E)={\tilde T}_{\beta\alpha}(E). \eqno (3)
$$
Moreover, since $V_\alpha=H-H_\alpha$ and $V_\beta=H-H_\beta$, the
relation~(3) holds true in the FBA case as well, that is, on the
energy shell the FBA {\it post-} and {\it prior-}amplitudes
coincide,
$$
\langle\phi_{\beta},\vec p_\beta|V_{\beta}|\phi_{\alpha},\vec
p_\alpha\rangle=\langle\phi_{\beta},\vec
p_\beta|V_{\alpha}|\phi_{\alpha},\vec p_\alpha\rangle.
$$

The above formulas are valid only in the case where colliding fragments do
not interact via long-range, Coulomb-like potentials at asymptotically
large separation distances. This can be formulated using the Zommerfeld
parameter of the channel
$$
\eta_{\gamma}=\sum_{i<j}\frac{Z^{(\gamma)}_{i}Z^{(\gamma)}_j}{v^{(\gamma)}_{ij}}
\qquad (\gamma=\alpha,\beta),
$$
where $Z^{(\gamma)}_{i}$ and $Z^{(\gamma)}_j$ are the total charges of the
colliding fragments $i$ and $j$, and $v^{(\gamma)}_{ij}$ is their relative
velocity. If the Zommerfeld parameter differs from zero, Eqs. (1) and (2)
become more complicated~\cite{Belkic,Shablov2010}, because the asymptotic
states $|\phi_{\alpha(\beta)},\vec p_{\alpha(\beta)}\rangle$ do not obey
the correct asymptotic conditions anymore.

Let us apply the above general formulas to the fast TI reaction
${\rm H}^+ +{\rm He}\to{\rm H}+{\rm He}^{2+}+e$ discussed in the
papers of Voitkiv \emph{et
al.}~\cite{Voitkiv_PRL,Voitkiv_JPB,Voitkiv_PRA}. The authors
utilize the {\it post-}amplitude, which in the nonsymmetrized FBA
can be written as
$$
T_{fi}^{FBA}(E)=\langle\phi_{p1},\varphi^-_{N2}(\vec k),\vec
p_H|V_{N1}+V_{p2}+V_{12}+V_{Np}|\Phi_0, \vec p_0\rangle. \eqno
(4a)
$$
In Eq.~(4a), electrons are labelled by ``1'' and ``2'', whereas ``$p$''
labels the fast proton projectile, and ``$N$'' the target nucleus. The
wave function $|\phi_{p1}\rangle$ is the bound (ground) state of atomic
hydrogen, $|\varphi_{N2}^-(\vec k)\rangle$ the continuum state of the
He$^+$ ion, $|\Phi_0\rangle$ the helium wave function, $\vec p_0$ the
proton momentum, $\vec p_H$ the hydrogen momentum, and $\vec k$ the
momentum of the emitted electron. This amplitude is equal to that in the
\emph{prior}-form
$$
{\tilde T}_{fi}^{FBA}(E)=\langle\phi_{p1},\varphi^-_{N2}(\vec k),\vec
p_H|V_{p1}+V_{p2}+V_{Np}|\Phi_0, \vec p_0\rangle. \eqno (4b)
$$

It should be noted that Eqs.~(1) and~(2) are applicable because there is
no long-range asymptotic interaction in the initial and final channels. In
the present case, it is clearly fulfilled in the initial channel (the He
atom is neutral, $Z_{He}=0$). It is also fulfilled in the final channel,
because $Z_{H}=0$, and we use in Eq.~(4) the spectral Coulomb functions
$\varphi^-_{N2}(\vec k)$ instead of plain waves, so that the neutral
hydrogen subsystem does not asymptotically interact with the He$^+$
subsystem.

From the equality of the FBA amplitudes~(4a) and~(4b) we find that
$$
\langle\phi_{p1},\varphi^-_{N2}(\vec k),\vec p_H|V_{p1}|\Phi_0,
\vec p_0\rangle=\langle\phi_{p1},\varphi^-_{N2}(\vec k),\vec
p_H|V_{N1}+V_{12}|\Phi_0, \vec p_0\rangle. \eqno (5)
$$
The matrix element on the left-hand side amounts to the OBK approximation.
It can be easily transformed into the overlap of the initial and final
wave functions described in Ref.~\cite{OBK}. The matrix element on the
right-hand side is the same OBK, but in the \emph{post}-form
representation. It is important to note that within FBA the physical
effect of the interaction of the transferred electron with the proton
projectile is exactly equal to that of the interaction of the same
electron with the residual target ion. This means that the $ee$-Auger
mechanism, which is attributed by Voitkiv \emph{et al.} to the $V_{12}$
contribution in the right-hand side of Eq. (5), is not independent and is
included in the OBK scenario. It has been repeatedly shown (see, for
instance, Ref.~\cite{Salim2010}) that, in the FBA {\it
prior}-amplitude~(4b), even the OBK term is not leading in some
kinematical situations. In other words, all four terms in~(4a) should be
considered in the general case.

\section{Distorted wave approximations \label{dwba}}
From Eqs.~(1) and (2) one can derive the higher Born terms as well as
different versions of the  distorted wave Born approximation (DWBA).
For example, in \cite{Belkic} the eikonal
approximation was derived, which introduces in the {\it prior}-form FBA matrix
element~(4b) a distorting phase factor,
$$
|\phi_{p1},\varphi^-_{N2}(\vec k),\vec p_H\rangle\to
e^{(i/v_p){\hat\delta_f}} |\phi_{p1},\varphi^-_{N2}(\vec k),\vec
p_H\rangle.
$$
The details concerning its derivation one can find in Ref.~\cite{Kim2012}.
We note that it is the asymptotic form of the product
$$
e^{(i/v_p){\hat\delta_f}}\to
\Lambda_f^{-}=\Lambda_{p2}^{-}\Lambda_{pN}^{-}\Lambda_{N1}^{-}\Lambda_{12}^{-},
\eqno (6)
$$
with
$$
\Lambda_{Z_1Z_2}^{-}=\exp\left(-\frac{\pi
Z_1Z_2}{2v_{rel}}\right)\Gamma\left(1-i\frac{Z_1Z_2}{v_{rel}}\right)
{_1}F_1\left[i\frac{Z_1Z_2}{v_{rel}},1; -i(v_{rel}r_{rel}+\vec v_{rel}\vec
r_{rel})\right],
$$
where $\vec{v}_{rel}$ and $\vec{r}_{rel}$ are the relative velocity and
position of the pair of particles. Each factor $\Lambda_{Z_1Z_2}^{-}$ in
(6) describes the distortion of interactions between different
constituents of the two final compound subsystems, H and He$^+$. In some
sense, it is a 4C model (in analogy with well known 3C and 6C models in
the scattering theory~\cite{berakdar03}).

The same procedure we can utilize in the case of the FBA matrix
element in the {\it post}-form, replacing in~(4a)
$$
|\Phi_0, \vec p_0\rangle\to e^{(i/v_p){\hat\delta_i}} |\Phi_0,
\vec p_0\rangle.
$$
Here, again,
$$
e^{(i/v_p){\hat\delta_i}}\to
\Lambda_i^{+}=\Lambda_{p1}^{+}\Lambda_{p2}^{+}\Lambda_{pN}^{+},
\eqno (7)
$$
and
$$
\Lambda_{Z_1Z_2}^{+}=\exp\left(-\frac{\pi
Z_1Z_2}{2v_p}\right)\Gamma\left(1+i\frac{Z_1Z_2}{v_p}\right)
{_1}F_1\left[-i\frac{Z_1Z_2}{v_p},1; i(v_pr_{rel}-\vec v_p\vec
r_{rel})\right].
$$
Each factor $\Lambda_{Z_1Z_2}^{+}$ in (7) describes the distortion due to
interactions between the projectile proton and different constituents of
the helium atom. This approximation is analogous to the 3C model.

Distortion factors (6) and (7) are typical of the continuum-distorted-wave
(CDW) model. The main requirement of this model is to obey the correct
Coulomb asymptotic conditions in the initial and final channels of the
reaction~\cite{Belkic}. These conditions are given by
($\gamma=\alpha,\beta$)~\cite{Dollard}
$$
e^{-iH_\gamma t}|\Psi^{\pm}_\gamma(\vec p_\gamma)\rangle\to
e^{-iE_\gamma t\pm i\eta_\gamma\ln|t|\pm iA_\gamma(\vec
p_\gamma)}|\phi_{\gamma},\vec p_\gamma\rangle, \qquad t\to
\mp\infty, \eqno (8)
$$
where $A_\gamma(\vec p_\gamma)$ is the so-called Dollard phase.
Representations (6) and (7) are not unique, and other forms are also
available (see, for instance, Refs.~\cite{Belkic,Crothers} and references
therein).

\section{$EE$-Auger mechanism in the rigorous scattering theory \label{results}}
In Refs.~\cite{Voitkiv_PRL,Voitkiv_JPB,Voitkiv_PRA}, the authors present
calculations of contributions from, as they suppose, different mechanisms.
These include the OBK (or capture-shakeoff), two-step (or independent
transfer ionization~\cite{Voitkiv_PRA}), $ee$-Thomas, and $ee$-Auger. The
details concerning calculation of these contributions, except that of the
$ee$-Auger, in Refs.~\cite{Voitkiv_PRL,Voitkiv_JPB,Voitkiv_PRA} are rather
scarce. Referring to the CDW model, Voitkiv \emph{et al.} use the
following formula for the $ee$-Auger amplitude (see Eq.~(7) of
Ref.~\cite{Voitkiv_PRA}):
$$
T_{fi}^{EEA}(E)=\langle\phi_{p1},\varphi^-_{N2}(\vec
k),\Lambda_{N1}^{-},\vec p_H|V_{12}|\Phi_0, \Lambda_{p1}^{+},\vec
p_0\rangle . \eqno (9)
$$
Thus, when summing the calculated in this way contribution with that of
the OBK, Voitkiv \emph{et al.} take into account the first-order
$ee$-Auger mechanism twice. Such a conclusion immediately follows from
Eq.~(5), which states that the $ee$-Auger contribution, connected with the
$V_{12}$ term on the right-hand side, is already taken into account in the
OBK approximation given by the left-hand side. And the presence of the
distorting factors does not principally change this apparent flaw in the
calculations of Refs.~\cite{Voitkiv_PRL,Voitkiv_JPB,Voitkiv_PRA}. It
should be also noted that formula (9) contradicts the CDW model. First, it
explicitly violates the correct asymptotic condition~(8) both in the
initial and in the final channels of the discussed reaction. Second, even
if one uses the correct asymptotic factors (6) and (7) in Eq.~(9), the
\emph{post}-form of the CDW model assumes that the perturbation is given
by the nonorthogonal kinetic energy $-\nabla_{N1}\cdot\nabla_{p1}$ (see
details in Refs.~\cite{Belkic,Crothers}, also in \cite{Ciappina}), which
is clearly not equivalent to $V_{12}$. It should be also remarked with
respect to Eq.~(9) that it violates not only the correct asymptotic
conditions, but also the identity of the helium electrons.

In Ref.~\cite{Voitkiv_PRA} the final distortion factor was neglected in
calculations, $\Lambda_f=1$, as being not so much significant. As remarked
in Ref.~\cite{Voitkiv_PRA}, without the initial distortion factor, i.e.,
when $\Lambda_i=1$, the contribution of the $ee$-Auger mechanism
calculated there becomes much larger, while that of $ee$-Thomas vanishes.
In view of these remarks, one might expect that neglecting the distortion
effects does not reduce the role of the $ee$-Auger mechanism. Thus, if
conclusions of Voitkiv~\emph{et al}. are correct, from Eq.~(5) it follows
that the discussed mechanism must manifest itself in the calculations
based on the OBK approximation, because the former is a part of the
latter.

The quantity that was studied numerically in
Refs.~\cite{Voitkiv_PRL,Voitkiv_JPB,Voitkiv_PRA} is the double
differential cross section (DDCS)
$$
\frac{d^2\sigma}{dk_{\perp} dk_z}=\frac{2k_{\perp}}{(2\pi)^5v_p^2}
\int\limits_0^{2\pi}d\varphi_k \int d^2{ q}_\perp|{T}_{fi}|^2, \eqno (10)
$$
which describes a 2D distribution of the momentum components of
the emitted electron ($k_x=k_\perp\cos\varphi_k,\
k_y=k_\perp\sin\varphi_k$). Numerical results for the DDCS using
the {\it prior}-OBK approximation (5) are shown in Fig. 1a. In
these calculations an accurate, highly correlated trial helium
function from \cite{Chuka06} is employed. The kinematical
situation is the same as that of Fig. 1 in
Ref.~\cite{Voitkiv_PRA}. We see a general tendency for the ejected
electron to be preferably emitted in the backward lobe ($k_z<0$),
which is typical for the OBK with highly correlated trial helium
wave functions. Next Born approximations are expected to
contribute to the forward lobe ($k_z>0$).

However, while there is a common feature such as a maximum located
at $k_z=0$, we find no maximum located at negative $k_z$ values
(approximately, at $k_z\approx-3.0$), in contrast to the results
presented in Fig. 1 of Ref.~\cite{Voitkiv_PRA}. The latter feature
is, according to Voitkiv \emph{et al.}, a clear signature of the
$ee$-Auger mechanism. Thus, our numerical calculations using the
accurate, highly correlated wave function of helium do not support
the findings of Refs.~\cite{Voitkiv_PRL,Voitkiv_JPB,Voitkiv_PRA}.
In this connection, it should be noted that angular correlations
due to the $V_{12}$ interaction play a very important role, if
being included in the trial helium wave function $\Phi (\vec r_1,
\vec r_2)$. They strongly influence the momentum distribution of
the emitted electron in the backward direction, which is due to
the shake-off mechanism~\cite{Schoeffler13}. And we see 
manifestation of their effect in the region $k_z<0$. However, it
is quite different from manifestation of the $ee$-Auger mechanism
claimed by Voitkiv \emph{et al.} According to Eq.~(5), the effect
of $V_{12}$ found in
Refs.~\cite{Voitkiv_PRL,Voitkiv_JPB,Voitkiv_PRA} (see Fig.1b) is
clearly cancelled by the other first-order mechanism, which
involves the target nucleus (the $V_{N1}$ term). This finding
markedly illustrates the importance of accounting for all the
binary interactions between the particles taking part in the
reaction.

Some comments should be made with regard to the equivalence of the
\emph{post}- and \emph{prior}-forms of the transition amplitude. It is
realized only if the exact helium wave function is employed. In that case,
we have the following equation for this function:
\begin{align*}
(\varepsilon_0^{He}-h_{10}-h_{20}-V_{N2})|\Phi_0\rangle
=(V_{N1}+V_{12})|\Phi_0\rangle. \tag{11}
\end{align*}
It can be readily shown, using energy conservation and properties of the
final asymptotic state, that the projectile-electron potential $V_{p1}$ in
the left-hand side of Eq.~(5) can be replaced with the operator
$(\varepsilon_0^{He}-h_{10}-h_{20}-V_{N2})$ occurring in the left-hand
side of Eq.~(11). Thus, the left- and right-hand sides of the
Schr\"odinger equation~(11) correspond to, respectively, the \emph{prior-}
and \emph{post-}matrix elements in Eq.~(5). This feature explains a well
documented fact that Eq.~(5) is fulfilled to a good approximation in the
case of an accurate trial helium function, which is typically obtained
from a variational procedure. But if the trial function is poor, then the
right-hand side of Eq.~(11), which is related to the \emph{post-}matrix
element in Eq.~(5), appears to yield a better approximation to the exact
result than in the case of the left-hand side. This observation explains
why the results of Voitkiv~\emph{et al}., using the correlated and
uncorrelated helium functions in their \emph{post}-amplitude, are similar
(see Fig.~7 of Ref.~\cite{Voitkiv_PRA}).


%
\section{Summary and conclusions \label{sum}}
We considered theoretically a transfer ionization channel in a fast
proton-helium collision, focusing on the so-called first-order
electron-electron capture mechanism proposed recently by Voitkiv \emph{et
al.}~\cite{Voitkiv_PRL,Voitkiv_JPB,Voitkiv_PRA}. It was shown, using
consistent quantum collision theory, that this specific mechanism is
included in a well known OBK scenario. The formula employed by
Voitkiv~\emph{et al}. for the transition amplitude is found to be
unjustified and contradicting the CDW model. The OBK calculations with an
accurate trial helium function exhibited no signature of the $ee$-Auger
process.

Neither distorted waves nor different representations of the amplitude
should change the basic physics of the process, which is essentially
governed by the projectile-target interaction. Voitkiv \emph{et al}. use
the \emph{post}-formulation, whereas many authors prefer the
\emph{prior}-formulation. However, in approximate treatments, one should
try to achieve their convergence (see, for instance,
Ref.~\cite{Ciappina06}), since the physics of the process does not depend
on the form of the matrix element. Interaction of the captured electron
with both nuclei is important. However, if it is accounted for within the
distorted-wave treatment, then such a treatment must be carried out in a
mathematically correct fashion. To our knowledge, all the requisites for
this problem can be found, for example, in the review article of Belki\'c
\emph{et al}~\cite{Belkic}.

\begin{acknowledgments}
This work is partially supported by the Russian Foundation for
Basic Research (Grant No. 11-01-00523-a). Calculations are
performed using the Moscow State University Research Computing
Center (supercomputer Chebyshev). The authors are grateful to O.
Chuluunbaatar and S. Houamer for their help. We are also grateful
to R. D\"orner and M. Sch\"offler for inspiring discussions.
\end{acknowledgments}

\bibliographystyle{unsrt}

\newpage

\begin{figure}
\centering
\includegraphics[width=\linewidth]{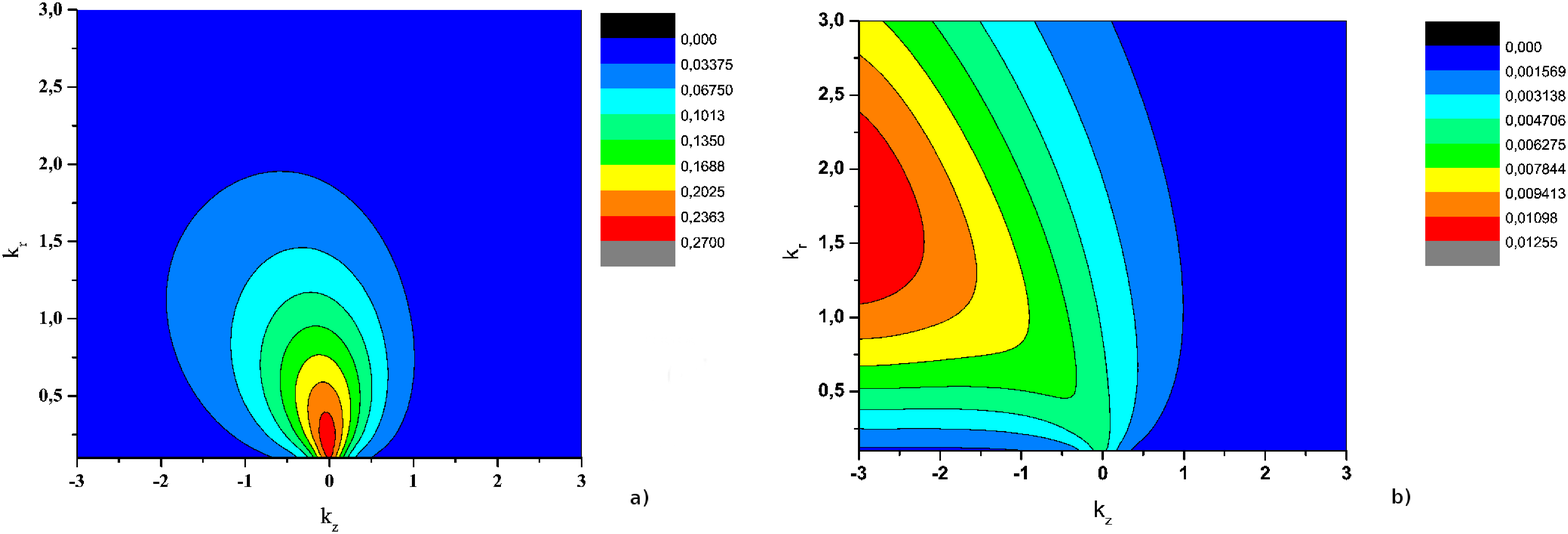}
\caption{(Color online) DDCS (10) in barn/(a.u.)$^2$,
$E_p=3.6$\,MeV. a) $T_{fi}$ = prior-OBK, b)
$T_{fi}=\langle\phi_{p1},\varphi^-_{N2}(\vec k),\vec
p_H|V_{12}|\Phi_0, \vec p_0\rangle$.}
\end{figure}

\end{document}